\documentclass[%
 aip,
 jmp,%
 amsmath,amssymb,
 reprint,%
]{revtex4-1}

\usepackage[cp1250]{inputenc}

\usepackage{amsthm}

\usepackage{bm}
\usepackage{dcolumn}

\newcommand{\proj}[2]{\left|#1\right>\left<#2\right|}

\newtheorem{theorem}{Theorem}[section]
\newtheorem{lemma}[theorem]{Lemma}

\newtheorem{remark}[theorem]{Remark}
\newtheorem{definition}[theorem]{Definition}

\newtheorem{proposition}[theorem]{Proposition}

\begin{document}
\title[Non-existence of partial transpose criterion analogues for three-by-three systems]{There is no direct generalization of positive partial transpose criterion to the three-by-three case}

\author{Łukasz Skowronek}
\email{lukasz.m.skowronek@gmail.com}
\affiliation{Brygadzistów 23A/3, 40-807 Katowice, Poland}
\altaffiliation[Former affiliation: ]{Instytut Fizyki im. Smoluchowskiego, Uniwersytet Jagielloński, Reymonta 4, 30-059 Kraków, Poland}
\date{\today}

\begin{abstract}
We show that there cannot exist a straightforward generalization of the famous positive partial transpose criterion to three-by-three systems. We call straightforward generalizations that use a finite set of positive maps and arbitrary local rotations of the tested two-partite state. In particular, we show that a family of extreme positive maps discussed in a paper by Ha and Kye \cite{HaKye11}, cannot be replaced by a finite set of witnesses in the task of entanglement detection in three-by-three systems. In a more mathematically elegant parlance, our result says that the convex cone of positive maps of the set of three-dimensional matrices into itself is not finitely generated as a mapping cone  
\end{abstract}
\maketitle
\section{Introduction}
In multi-partite quantum systems, separable states are a generalization of pure product states to density matrices. They are defined as convex combinations of one-dimensional projections onto product states, 
\begin{equation}
\rho = \sum_{i=1}^k\alpha_i\proj{\phi^1_i}{\phi^1_i}\otimes\ldots\otimes\proj{\phi^n_i}{\phi^n_i},
\end{equation}
where $\sum_{i=1}^k\alpha_i=1$ and $\forall_{i}\alpha_i>0$.

Separable state exhibit fully classical correlations between the individual subsystems. Only states that are not separable may therefore be useful for tasks like quantum cryptography \cite{Bennett84,GisinRevModPhys2002}. Such non-separable states are called \emph{entangled}. In the two-partite case, all entangled states have been shown to be capable of violating a Bell type inequality in a setting where separable states are useless \cite{Masanes2006,Masanes2008,MasanesLiangDoherty2008}. On the other hand, correlations obtained for an entangled state may exhibit a hidden variable description \cite{Werner89}. Unlike for pure product states, it turns out to be very difficult to test whether a given density matrix is separable or entanged. A notable exception to this rule are two-partite density matrices for $2\times 2$ and $2\times 3$ systems, which turn out to be separable if and only if 
\begin{equation}
\left(\mathsf{id}\otimes\mathsf{t}\right)\rho\geqslant 0,
\label{PPTCriterion}
\end{equation}
where \textsf{id} and \textsf{t} stand for the identity and the transpose maps, respectively, and $\rho$ denotes the density matrix in question. This is the famous positive partial transpose (PPT) criterion by Peres and Horodecki \cite{Peres96,Horodeccy96}, which builds on the decomposability of positive maps of $2\times 2$ matrices into $2\times 2$ or $3\times 3$ matrices \cite{Stormer63,Woronowicz76}.

It is well known that the PPT test is not a sufficient separability criterion for $3\times 3$ systems. In the $3\times 3$ case, there exist entangled states with positive partial transpose \cite{Horodecki97}. They are examples of so-called \emph{bound entangled states} .

In the present paper, we shall prove that no generalization of \eqref{PPTCriterion} of the form
\begin{equation}
\left(\mathsf{id}\otimes\Phi_i\right)\left(A\otimes B\right)\rho\left(A\otimes B\right)^{\ast}\geqslant 0\,\forall_{i=1,2,\ldots,m;A,B\in M_{3}}
\label{eqCritFamily}
\end{equation}
can be a sufficient criterion for separability in $3\times 3$ systems. Here, $\left\{\Phi_i\right\}_{i=1}^k$ stands for a finite set of positive maps and $A$, $B$ denote arbitrary $3\times 3$ matrices with complex entries. We denote the set of all such matrices by $M_3$. We immediately see that the similarity transform by $A$ is superfluous in \eqref{eqCritFamily} and we can therefore replace \eqref{eqCritFamily} with
\begin{equation}\label{eqCritFamily2}
\left(\mathsf{id}\otimes\Phi_i\right)\left(\mathbb{I}\otimes B\right)\rho\left(\mathbb{I}\otimes B\right)^{\ast}\geqslant 0\,\forall_{i=1,2,\ldots,m;B\in M_{3}}
\end{equation}
We claim that no such criterion can be sufficient for separability in $3\times 3$ systems. This result has been obtained by the author in spring 2011 and presented at a seminar at ICFO on 2011/05/09. Due to difficult personal conditions and taking up a job outside academia, the result has not been available in a polished written up form until now. 

\section{Proof}
Let us assume that $\Phi$ is an element of the set $\mathcal{M}_{3,3}$ of linear maps of the set of three-dimensional matrices $M_3$ into itself. It will be convenient for us to use the Jamiołkowski-Choi transform \cite{Jamiolkowski72,Choi75} of $\Phi$,
\begin{equation}\label{eqChoi}
C_{\Phi} = \sum_{i,j=1,2,3}e_{ij}\otimes\Phi\left(e_{ij}\right),
\end{equation}
where $e_{ij}$ are elements of a canonical basis of $M_3$. The object $C_{\Phi}$ is called the \textit{Choi matrix} of $\Phi$ after \cite{Choi75}, and the mapping $J:\mathcal{M}_{3,3}\ni\Phi\mapsto C_{\Phi}\in M_9$ is an isomorphism between the space of maps of the set of three-dimensional matrices into itself and the set of nine-dimensional matrices. This isomorphism is named the \textit{Jamiołkowski isomorphism} after \cite{Jamiolkowski72}. Interestingly, if we define a natural inner product in $\mathcal{M}_{3,3}$,
\begin{equation}
\left<\Phi,\Psi\right>'':=\sum_{i,j=1,2,3}\mathrm{Tr}\left(\Phi\left(e_{ij}\right)\Psi\left(e_{ij}\right)^{\ast}\right),
\end{equation} 
$J$ also turns out to be an isometry,
\begin{equation}\label{eqIso}
\left<\Phi,\Psi\right>''=\mathrm{Tr}\left(C_{\Phi}C_{\Psi}^{\ast}\right)=:\left<C_{\Phi},C_{\Psi}\right>',
\end{equation}
where the conjugation is taken with the assumption of orthonormality of the canonical basis of $\mathbb{C}^3$ underlying $e_{ij}$. The inner product $\left<.,.\right>'$ is simply a Hilbert-Schmidt product in $M_9$. Similarly, $\left<.,.\right>''$ is a Hilbert-Schmidt product in $\mathcal{M}_{3,3}$.

We shall need the following properties of Jamiołkowski isomorphism and the Hilbert-Schmidt product in $\mathcal{M}_{3,3}$ (cf. e.g. \cite{Skowronek11}).
\begin{proposition}\label{propProj}
Let $A=\sum_{i,j=1,2,3}A_{ij}e_{ij}$ be an element of $M_3$ and let $\mathsf{Ad}_A:M_3\ni X\mapsto AXA^{\ast}\in M_3$ be the conjugation map by $A$. Then we have
\begin{equation}
J\left(\mathsf{Ad}_A\right)=C_{\mathsf{Ad}_A}=\left|\alpha\right>\left<\alpha\right|,
\end{equation}
where $\alpha:=\sum_{i,j}A_{ij}e_i\otimes e_j$ for $\left\{e_i\right\}_{i=1}^3$ a canonical basis of $\mathbb{C}^3$ corresponding to the canonical basis $\left\{e_{ij}\right\}_{i,j=1,2,3}$ of $M_3$. 
\end{proposition}
In other words, arbitrary conjugation maps in $M_3$ map to arbitrary non-normalized one-dimensional projections in $\mathbb{C}^3$.
\begin{proposition}\label{propAst}
For all $\Phi,\Sigma,\Psi\in\mathcal{M}_{3,3}$, we have
\begin{equation}
\left<\Phi\circ\Sigma,\Psi\right>''=\left<\Sigma,\Phi^{\ast}\circ\Psi\right>''
\end{equation} 
\end{proposition}

\begin{proposition}\label{propAstAst}
For all $\Theta,\Phi,\Sigma,\Psi\in\mathcal{M}_{3,3}$, we have
\begin{equation}
\left<\Theta\circ\Phi\circ\Sigma,\Psi\right>''=\left<\Phi,\Theta^{\ast}\circ\Psi\circ\Sigma^{\ast}\right>''
\end{equation} 
\end{proposition}
We also have the trivial
\begin{proposition}\label{propComp}
For arbitrary $\Phi,\Psi\in\mathcal{M}_{3,3}$
\begin{equation}
C_{\Phi\circ\Psi}=\left(\mathbb{I}\otimes\Phi\right)C_{\Psi}
\end{equation}
\end{proposition}
and the slightly more involved
\begin{proposition}\label{propAdAB}
For arbitrary $A,B\in M_3$ and $\Phi\in\mathcal{M}_{3,3}$, we have
\begin{equation}
C_{\mathsf{Ad}_A\circ\Phi\circ\mathsf{Ad}_B}=\mathsf{Ad}_{B^t\otimes A}C_{\Phi}
\end{equation}
\end{proposition}
which we prove in Appendix \ref{appAdAB}.

Let us note that the convex cone $\mathcal{P}_{3,3}$ of positive (i.e. positivity preserving) maps from $M_3$ into $M_3$ is a subset of the set $\mathcal{H}_{3,3}$ of Hermiticity preserving maps (cf. e.g. \cite{Pillis67}). It is easy to see that $\mathcal{H}_{3,3}$ is a linear space over $\mathbb{R}$ and $\left<.,.\right>''$ induces a symmetric inner product in $\mathcal{H}_{3,3}$. Thus for every convex cone $\mathcal{K}\subset\mathcal{H}_{3,3}$, we can define the \textit{dual cone} of $\mathcal{K}$,
\begin{equation}
\mathcal{K}^{\circ}:=\left\{\Psi\in\mathcal{H}_{3,3}|\left<\Psi,\Phi\right>''\geqslant 0\forall_{\Phi\in\mathcal{K}}\right\}
\end{equation}
Among the convex cones contained in $\mathcal{H}_{3,3}$, and more generally in $\mathcal{H}_{m,n}$ with $m$, $n$ arbitrary, there is a distinguished class of so-called \textit{mapping cones}, introduced by St\o rmer \cite{Stormer86}.
\begin{definition}\label{defMap}
Let $\mathcal{K}\subset\mathcal{P}_{3,3}$ be a cone. We call $\mathcal{K}$ a \textit{mapping cone} if $\mathcal{K}$ is closed, not consisting only of the zero map, and
\begin{equation}
\mathsf{Ad}_A\circ\Phi\circ\mathsf{Ad}_B\in\mathcal{K}\forall_{\Phi\in\mathcal{K}}\forall_{A,B\in M_3}
\end{equation}
\end{definition}
In other words, $\mathcal{K}$ is closed with respect to left and right multiplications by arbitrary conjugation maps. Note that convexity is not explicitly assumed in Definition \ref{defMap}, however we shall always assume $\mathcal{K}$ to be convex in what follows. The papers \cite{Stormer08,Stormer09,Stormer09II,SSZ09,Skowronek11,JohnstonStormer12,JSS12} show several interesting properties of mapping cones in the finite-dimensional case, specifically concerning their duals and their relation to operator algebras. The cone of positive maps $\mathcal{P}_{m,n}$ is obviously the biggest mapping cone in $\mathcal{H}_{m,n}$, whereas its dual $\mathcal{SP}_{m,n}$, called the set of \textit{superpositive maps} \cite{Ando04}, is in a sense the smallest one, because it is contained in all convex mapping cones. As long as convex cones are concerned, which is the case in this paper and in \cite{Skowronek11}, we can rightly call $\mathcal{SP}_{m,n}$ the smallest mapping cone. It turns out that (cf. e.g. \cite{SSZ09}) $\mathcal{SP}_{m,n}$ can be characterized as the set of convex combinations of the form
\begin{equation}
\sum_{i=1}^n\mathsf{Ad}_{A_i},\,\mathrm{rk}\left(A_i\right)=1
\end{equation}  
where $\mathrm{rk}\left(A\right)$ denotes the rank of a matrix $A$. From Proposition \ref{propProj}, we easily see that, up to scaling, superpositive maps are in a one-to-one correspondence with separable states via Jamiołkowski isomorphism. It is also not difficult to deduce (for more details cf. \cite{Jamiolkowski72,SSZ09}) that the set of all positive maps $\Phi$ corresponds via Jamiołkowski isomorphism to the set of \textit{block positive} Choi matrices, i.e. Choi matrices $C_{\Phi}$ such that
\begin{equation}\label{eqBP}
\left<x\otimes y\right|C_{\Phi}\left|x\otimes y\right>>0\forall_{x,y\in\mathbb{C}^3}.
\end{equation}

Let us denote the set of separable states in $3\times 3$ systems by $\mathsf{Sep}_{3\times 3}$. All the above facts give us enough information to prove the following equivalence.
\begin{proposition}\label{propEquivConvhull}
The nonexistence of a separability criterion of the form
\begin{equation}\label{eqCritFamily3}
\rho\in\mathsf{Sep}_{3\times 3}\Leftrightarrow\forall_{i=1,2,\ldots,k}\forall_{B\in M_3}\left(\mathbb{I}\otimes\Phi_i\right)\mathsf{Ad}_{\mathbb{I}\otimes B}\rho\geqslant 0
\end{equation}
is equivalent to the fact that there does not exist a finite set of positive maps $\left\{\Xi_i\right\}_{i=1}^k\in\mathcal{P}_{3,3}$ such that
\begin{equation}\label{eqConvhull}
\mathcal{P}_{3,3}=\overline{\mathrm{convhull}\left(\left\{\mathsf{Ad}_X\circ\Xi_i\circ\mathsf{Ad}_Y|i;X,Y\in M_3\right\}\right)},
\end{equation}
where $\mathrm{convhull}$ denotes the convex hull of a set and the overline refers to topological closure. 
\begin{proof}
Let us denote by $\Psi$ the inverse Jamiołkowski transform of a density matrix $\rho\in M_9$. By Propositions \ref{propProj}--\ref{propAdAB} and the fact that Jamiołkowski isomorphism is an isometry, we have
\begin{multline}\label{eqEquiv}
\left(\mathbb{I}\otimes\Phi_i\right)\mathsf{Ad}_{\mathbb{I}\otimes B}\rho\geqslant 0\Leftrightarrow
C_{\Phi_i\circ\mathsf{Ad}_B\circ\Psi}\geqslant 0\Leftrightarrow\\
\Leftrightarrow\left<\left|\alpha\right>\left<\alpha\right|,C_{\Phi_i\circ\mathsf{Ad}_B\circ\Psi}\right>'\geqslant 0\forall_{\alpha\in\mathbb{C}^9}\Leftrightarrow\\\Leftrightarrow\left<C_{\mathsf{Ad}_A},C_{\Phi_i\circ\mathsf{Ad}_B\circ\Psi}\right>'\geqslant 0\forall_{A\in M_3}\Leftrightarrow\\\Leftrightarrow\left<\mathsf{Ad}_A,\Phi_i\circ\mathsf{Ad}_B\circ\Psi\right>''\geqslant 0\forall_{A\in M_3}\Leftrightarrow\\\Leftrightarrow\left<\mathsf{Ad}_{B^{\ast}}\circ\Phi^{\ast}_i\circ\mathsf{Ad}_A,\Psi\right>''\geqslant 0\forall_{A\in M_3}
\end{multline} 
Let us assume that \eqref{eqCritFamily3} is a sufficient criterion for separability of a state $\rho$. If we additionally assume that \eqref{eqConvhull} does not hold for $\Xi_i=\Phi^{\ast}_i$, $i=1,2,\ldots,k$, then by the identity (cf. \cite{Rockafellar} and a comment in \cite{Skowronek11})
\begin{equation}\label{eqCircCirc}
\mathcal{K}^{\circ\circ}=\mathcal{K}
\end{equation}
applied to the closure of the convex hull of maps of the form $\mathsf{Ad}_{B^{\ast}}\circ\Phi_i^{\ast}\circ\mathsf{Ad}_A$, $A,B\in M_3$, we see that there must exist an element $\Psi$ of $\mathcal{P}_{3,3}\setminus\mathcal{SP}_{3,3}$ such that
\begin{equation}\label{eqGeExcept}
\left<\mathsf{Ad}_{B^{\ast}}\circ\Phi_i^{\ast}\circ\mathsf{Ad}_A,\Psi\right>''\geqslant 0\forall_{i;A,B\in M_3}
\end{equation}
We already know by \eqref{eqEquiv} that \eqref{eqGeExcept} is equivalent to 
\begin{equation}\label{eqGeExcept2}
\left(\mathbb{I}\otimes\Phi_i\right)\mathsf{Ad}_{\mathbb{I}\otimes B}C_{\Psi}\geqslant 0\forall_{i;B\in M_3}
\end{equation}
this implies, up to scaling, $C_{\Psi}\in\mathsf{Sep}_{3,3}$, which is a contradiction since $J^{-1}\left(\mathsf{Sep}_{3,3}\right)\subset\mathcal{SP}_{3,3}$, where the last inclusion turns into an equality if we allow arbitrary scaling of the separable density matrices by non-negative factors. We are led to a conclusion that \eqref{eqCritFamily3} cannot be a sufficient criterion for separability. If it is, then \eqref{eqConvhull} must hold. Conversely, if the cone of positive maps can be written in the form \eqref{eqConvhull}, then by the property \eqref{eqCircCirc} applied to $\mathcal{K}=\mathcal{P}_{3,3}$ ($\mathcal{K}^{\circ}=\mathcal{SP}_{3,3}$), a sufficient criterion for a map $\Psi$ to be superpositive, or in other words, for the Choi matrix $C_{\Psi}$ to be a non-negative multiple of a separable density matrix, is \eqref{eqGeExcept} with $\Xi_i$ substituted for $\Phi_i^{\ast}$, $X$ substituted for $B^{\ast}$ and $Y$ substituted for $B$. That is further equivalent to Equation \eqref{eqGeExcept2} with $\Xi_i^{\ast}$ in place of $\Phi_i$ and $X^{\ast}$ substituted for $B$. Up to relabeling, this is the same as \eqref{eqCritFamily3}, which finishes the proof.
\end{proof}
\end{proposition} 
The main result of this paper is therefore equivalent to the fact that does not exist a finite set of generators $\left\{\Xi_i\right\}_{i=1}^k$ of $\mathcal{P}_{3,3}$ considered as a mapping cone. Let us assume for a moment that such set of generators exists. For $C\in M_9$, let us also denote by $\left<.\otimes y\right|C\left|.\otimes y\right>$ and $\left<x\otimes .\right|C\left|x\otimes .\right>$ matrices in $M_3$ with matrix elements $\sum_{k,l=1}^3\bar{y}^kC_{ik,jl}y^l$ and $\sum_{k,l=1}^3\bar{x}^kC_{ki,lj}x^l$, respectively. We have the following.
\begin{proposition}\label{propPoints}
Let $\left\{\Xi_i\right\}_{i=1}^k$ be a set of positive maps in $\mathcal{P}_{3,3}$ such that equality \eqref{eqConvhull} holds. Then we can assume that:
\begin{enumerate}
	\item All $\Xi_i$'s are extreme.
	\item $\Xi_i=\mathsf{Ad}_X\circ\Xi_j\circ\mathsf{Ad}_Y$ does not hold for any $i,j\in\left\{1,2,\ldots,k\right\}$, $X,Y\in M_3$ unless $i=j$.
	\item For any $\Phi$ positive, extreme and such that the matrices $\left<.\otimes y\right|C_{\Phi}\left|.\otimes y\right>$ and $\left<x\otimes .\right|C_{\Phi}\left|x\otimes .\right>$ are of rank at least $2$ for all $x,y\in\mathbb{C}^3$, $\Phi=\mathsf{Ad}_X\circ\Xi_i\circ\mathsf{Ad}_Y$ for some $i\in\left\{1,2,\ldots,k\right\}$ and $X,Y\in M_3$.
	\item If $\Phi$ positive and extreme is not of the form $\Phi = \mathsf{Ad}_Z$ or $\Phi = \mathsf{Ad}_Z\circ\mathsf{t}$  for some $Z\in M_3$, i.e. $\Phi$ is not decomposable, the matrices $X$ and $Y$ in $\Phi=\mathsf{Ad}_X\circ\Xi_i\circ\mathsf{Ad}_Y$ must be invertible. 
	\item For any $\Phi$ positive, extreme and such that the matrices $\left<.\otimes y\right|C_{\Phi}\left|.\otimes y\right>$ and $\left<x\otimes .\right|C_{\Phi}\left|x\otimes .\right>$ are of rank at least $2$ for all $x,y\in\mathbb{C}^3$, there is exactly one $i$ such that $\Phi=\mathsf{Ad}_X\circ\Xi_i\circ\mathsf{Ad}_Y$ for some $X,Y\in M_3$.
\end{enumerate}
\begin{proof}
Points 1) and 2) are obvious (in particular, point 2) simply states the fact that we can keep removing some $\Xi_i$'s until none of them are redundant as generators of the mapping cone \eqref{eqConvhull}). Point 3) needs a longer proof that we provide in Appendix \ref{appClosed}. Point 4) is a consequence of the fact that positive maps in $2\times 2$ and $2\times 3$ systems are \textit{decomposable}, i.e. they can be written in the form $\sum_i\mathsf{Ad}_{X_i}+\sum_j\mathsf{Ad}_{X'_i}\circ\mathsf{t}$ for some $X_i,X'_i\in M_3$ \cite{Stormer63,Woronowicz76}. Non-invertible $A$ or $B$ would bring us back to this basic case, as it is easily seen from Proposition \ref{propAdAB}. Point 5) follows because from point 3) we already know that there must exist $X,Y\in M_3$ and $i\in\left\{1,2,\ldots,k\right\}$ such that equality $\Phi=\mathsf{Ad}_X\circ\Xi_i\circ\mathsf{Ad}_Y$ holds. Moreover, by point 4) the matrices $X$ and $Y$ must be invertible. If $i$ were not unique, we would have $\Phi=\mathsf{Ad}_Z\circ\Xi_j\circ\mathsf{Ad}_T=\mathsf{Ad}_X\circ\Xi_i\circ\mathsf{Ad}_Y$ for some $j\neq i$ and $Z,T\in M_3$ invertible. This would imply $\Xi_j=\mathsf{Ad}_{Z^{-1}X}\circ\Xi_i\circ\mathsf{Ad}_{YT^{-1}}$, contradicting point 2). 
\end{proof}
\end{proposition}
It is easy to see from the above proposition that the existence of an infinite family $\left\{\Phi_{\alpha}\right\}_{\alpha\in\mathcal{A}}\subset\mathcal{P}_{3,3}$ of positive, indecomposable, extreme maps such that $\mathrm{rk}\left(\left<x\otimes .\right|C_{\Phi_{\alpha}}\left|x\otimes .\right>\right)\geqslant 2$ and $\mathrm{rk}\left(\left<.\otimes y\right|C_{\Phi_{\alpha}}\left|.\otimes y\right>\right)\geqslant 2$ for all $x,y\in\mathbb{C}^3$ and
\begin{equation}\label{eqNexists}
\neg(\exists_{\alpha_1\neq\alpha_2}\exists_{X,Y\in M_3}\Phi_{\alpha_1}=\mathsf{Ad}_X\circ\Phi_{\alpha_2}\circ\mathsf{Ad}_Y)
\end{equation}  
would preclude the existence of a finite set of mapping cone generators of $\mathcal{P}_{3,3}$. It turns out that the family of extreme, indecomposable positive maps
\begin{widetext}
\begin{equation}\label{eqHaKyeFamily}
\Phi_{t}:M_3\ni\left[x_{ij}\right]\mapsto\left[
\begin{array}{ccc}
a_tx_{11}+b_tx_{22}+c_tx_{33}&-x_{12}&-x_{13}\\
-x_{21}&c_tx_{11}+a_tx_{22}+b_tx_{33}&-x_{23}\\
-x_{31}&-x_{32}&b_tx_{11}+c_tx_{22}+a_tx_{33}
\end{array}\right]\in M_3
\end{equation}
\end{widetext}
with $a_t=(1-t)^2/(1-t+t^2)$, $b_t=t^2/(1-t+t^2)$, $c_t=1/(1-t+t^2)$, $t\in\left[0,1\right)$,
introduced by Ha and Kye \cite{HaKye11}, has the required properties. To show this, let us first prove the following facts about of $\Phi_{t}$
\begin{proposition}\label{propPoints2}
Let $\Phi_{t}$ denote positive maps of the form \eqref{eqHaKyeFamily}. Let $x=\left[x^1,x^2,x^3\right]$ and $y=\left[y^1,y^2,y^3\right]$ denote normalized vectors in $\mathbb{C}^3$. Then
\begin{enumerate}
\item $\mathrm{det}\left(\left<.\otimes y\right|C_{\Phi_{t}}\left|.\otimes y\right>\right)$ only depends on $\left|y^1\right|$, $\left|y^2\right|$ and $\left|y^3\right|$
\item $\left<.\otimes y\right|C_{\Phi_{t}}\left|.\otimes y\right>$ is singular if and only if
\begin{equation}\label{eqnorms1}
\left|y^1\right|=\left|y^2\right|=\left|y^3\right|=\frac{1}{\sqrt{3}}
\end{equation}
or
\begin{equation}\label{eqnorms2}
\left|y^1\right|=0,\,\left|y^2\right|=\sqrt{\frac{t}{1+t}},\,\left|y^3\right|=\sqrt{\frac{1}{1+t}}
\end{equation}
or
\begin{equation}\label{eqnorms3}
\left|y^1\right|=\sqrt{\frac{1}{1+t}},\,\left|y^2\right|=0,\,\left|y^3\right|=\sqrt{\frac{t}{1+t}}
\end{equation}
or
\begin{equation}\label{eqnorms4}
\left|y^1\right|=\sqrt{\frac{t}{1+t}},\,\left|y^2\right|=\sqrt{\frac{1}{1+t}},\,\left|y^3\right|=0
\end{equation}
\item $\left<.\otimes y\right|C_{\Phi_{t}}\left|.\otimes y\right>$ is never of rank $1$ or $0$
\item For every $y$ that satisfies one of the condition sets \eqref{eqnorms1}-\eqref{eqnorms4}, there is, up to multiplication by a constant factor, exactly one $x$ such that
\begin{equation}\label{eqxyzero}
\left<x\otimes y\right|C_{\Phi_t}\left|x\otimes y\right>=0
\end{equation} 
holds
\item For $y_1$ and $y_2$, $y_1\neq y_2$, substituted in place of $y$ in \eqref{eqxyzero}, the corresponding vectors $x$ are never multiples of each other
\item $\left<x\otimes .\right|C_{\Phi_{t}}\left|x\otimes .\right>$ is never of rank $1$ or $0$
\end{enumerate}
\begin{proof}
The Choi matrix of $\Phi_t$ equals
\begin{equation}\label{eqChoiPhit}
\left[\begin{array}{ccccccccc}
a_t&0&0&0&-1&0&0&0&-1\\
0&c_t&0&0&0&0&0&0&0\\
0&0&b_t&0&0&0&0&0&0\\
0&0&0&b_t&0&0&0&0&0\\
-1&0&0&0&a_t&0&0&0&-1\\
0&0&0&0&0&c_t&0&0&0\\
0&0&0&0&0&0&c_t&0&0\\
0&0&0&0&0&0&0&b_t&0\\
-1&0&0&0&-1&0&0&0&a_t\\
\end{array}\right]
\end{equation}
By an elementary calculation, one can show that
\begin{widetext} 
\begin{equation}\label{eqContrY}
\left<.\otimes y\right|C_{\Phi_t}\left|.\otimes y\right>=\left[\begin{array}{ccc}
a_t\left|y^1\right|^2+b_t\left|y^2\right|^2+c_t\left|y^3\right|^2&-\bar y^1y^2&-\bar y^1y^3\\
-y^1\bar y^2&c_t\left|y^1\right|^2+a_t\left|y^2\right|^2+b_t\left|y^3\right|^2&-\bar y^2y^3\\
-y^1\bar y^3&-y^2\bar y^3&b_t\left|y^1\right|^2+c_t\left|y^2\right|^2+a_t\left|y^3\right|^2
\end{array}
\right]
\end{equation}
\end{widetext}
It is now almost obvious that the determinant of the matrix in \eqref{eqContrY} is dependent only on the moduli of $y^1$, $y^2$ and $y^3$. Concretely, we have
\begin{multline}\label{eqFt}
F_t\left(y\right):=\mathrm{det}\left(\left<.\otimes y\right|C_{\Phi_t}\left|.\otimes y\right>\right)=\\
=a_tb_tc_t\left(\left|y^1\right|^6+\left|y^2\right|^6+\left|y^2\right|^6\right)+\\
+\left(a_tb_t^2+b_tc_t^2+c_ta_t^2-c_t\right)\left(\left|y^2\right|^2\left|y^3\right|^4+\left|y^3\right|^2\left|y^1\right|^4+\right.\\+\left.\left|y^1\right|^2\left|y^2\right|^4\right)+\left(a_tc_t^2+b_ta_t^2+c_tb_t^2-b_t\right)\cdot\\
\cdot\left(\left|y^1\right|^2\left|y^3\right|^4+\left|y^2\right|^2\left|y^1\right|^4+\left|y^3\right|^2\left|y^2\right|^4\right)+\\
+\left(a_t^3+b_t^3+c_t^3+3a_tb_tc_t-2\right)\left|y^1\right|^2\left|y^2\right|^2\left|y^3\right|^2
\end{multline}
where we introduced the notation $F_t\left(y\right)$ for later convenience. From formula \eqref{eqFt}, we see that point 1) of the proposition holds. If we note that $a_t+b_t+c_t=2\,\forall_{t\in\left[0,1\right)}$ and that the matrix
\begin{equation}
\left[\begin{array}{ccc}
2&-1&-1\\
-1&2&-1\\
-1&-1&2
\end{array}\right]
\end{equation}
is singular, we immediately see that $y^1=y^2=y^3=1/\sqrt{3}$ solves $F_t\left(y\right)=0$. Because $F_t\left(y\right)$ only depends on the moduli of the coordinates of $y$, we see that all vectors $y$ that fulfill equation \eqref{eqnorms1} correspond to singular $\left<.\otimes y\right|C_{\Phi_t}\left|.\otimes y\right>$. Indeed, for such vectors the matrix $\left<.\otimes y\right|C_{\Phi_t}\left|.\otimes y\right>$ is a multiple of
\begin{equation}\label{eqPartConjY}
\left[\begin{array}{ccc}
2&-e^{i\left(\phi_2-\phi_1\right)}&-e^{i\left(\phi_3-\phi_1\right)}\\
-e^{i\left(\phi_1-\phi_2\right)}&2&-e^{i\left(\phi_3-\phi_2\right)}\\
-e^{i\left(\phi_1-\phi_3\right)}&-e^{i\left(\phi_2-\phi_3\right)}&2
\end{array}\right],
\end{equation}
where $y^1=\left|y^1\right|e^{i\phi_1}$, $y^2=\left|y^2\right|e^{i\phi_2}$ and $y^3=\left|y^3\right|e^{i\phi_3}$. This matrix is singular.

We now show that there are no other $y$ with non-vanishing $y^1$, $y^2$ and $y^3$ such that $F_t\left(y\right)=0$. Because $C_{\Phi_t}$ is block positive (cf. eq. \eqref{eqBP}), points at which $F_t\left(y\right)$ vanishes must be minima of $F_t$, considered as a function of the coordinates of $y$. More conveniently, if we define $l_1:=\left|y^1\right|^2$, $l_2:=\left|y^2\right|^2$ and $l_3:=\left|y^3\right|^2$ and write $F_t\left(y\right)$ as a function $F_t\left(l_1,l_2,l_3\right)$ of $l_1$, $l_2$ and $l_3$, singular $C_{\Phi_t}$ will correspond to minima of $F_t\left(l_1,l_2,l_3\right)$, as long as we are only interested in $y$'s with all coordinates non-vanishing (i.e. $l_i>0\,\forall_{i=1,2,3}$). The partial derivatives of $F_t\left(l_1,l_2,l_3\right)$ with respect to $l_1$, $l_2$ and $l_3$ equal
\begin{eqnarray}
3A_tl_1^2+B_t\left(2l_1l_3+l_2^2\right)+C_t\left(2l_1l_2+l_3^2\right)+D_tl_2l_3\label{eqPart1}\\
3A_tl_2^2+B_t\left(2l_2l_1+l_3^2\right)+C_t\left(2l_2l_3+l_1^2\right)+D_tl_3l_1\label{eqPart2}\\
3A_tl_3^2+B_t\left(2l_3l_2+l_1^2\right)+C_t\left(2l_3l_1+l_2^2\right)+D_tl_1l_2\label{eqPart3}
\end{eqnarray}
respectively, where we have introduced the notation
\begin{eqnarray}
A_t&:=&a_tb_tc_t\\
B_t&:=&a_tb^2_t+b_tc_t^2+c_ta^2_t-c_t\\
C_t&:=&a_tc_t^2+b_ta_t^2+c_tb_t^2-b_t\\
D_t&:=&a_t^3+b_t^3+c_t^3+3a_tb_tc_t-3a_t-2
\end{eqnarray}
By equating the partial derivatives \eqref{eqPart1}-\eqref{eqPart3} to zero and summing up the resulting equations, we get
\begin{multline}\label{eqThreeDerivs}
\left(3A_t+B_t+C_t\right)\left(l_1^2+l_2^2+l_3^2\right)+\\+\left(2B_t+2C_t+D_t\right)\left(l_1l_2+l_2l_3+l_3l_1\right)=0
\end{multline}
which must be fulfilled by squared moduli of coordinates of all vectors $y$ such that $\mathrm{det}\left(\left<.\otimes y\right|C_{\Phi_{t}}\left|.\otimes y\right>\right)=0$. Because we already know that $l_1=l_2=l_3$ results in a singular $\left<.\otimes y\right|C_{\Phi_{t}}\left|.\otimes y\right>$, we must have
\begin{equation}
3A_t+B_t+C_t = -2B_t-2C_t-D_t
\end{equation}
If we can prove that the above expressions are non-vanishing for all $t$, equation \eqref{eqThreeDerivs} will imply
\begin{equation}
l_1^2+l_2^2+l_3^2=l_1l_2+l_2l_3+l_3l_1
\end{equation}
which by Schwartz inequality can only hold if $l_1=l_2=l_3$. The expression $3A_t+B_t+C_t$ indeed does not vanish for any $t\in\left[0,1\right)$. Elementary algebra shows that
\begin{multline}\label{eqExprABC}
3A_t+B_t+C_t=\left(a_t+b_t+c_t\right)\left(a_tb_t+b_tc_t+c_ta_t\right)+\\-c_t-b_t=2\left(a_tb_t+b_tc_t+c_ta_t\right)-c_t-b_t=\\=\frac{\left(1-t\right)^3}{\left(1-t+t^2\right)^2}
\end{multline}
which is positive for all $t\in\left[0,1\right)$. In this way we have proved that the only choice of $l_1\neq 0$, $l_2\neq 0$, $l_3\neq 0$ such that $F_t\left(l_1,l_2,l_3\right)=0$ is $l_1=l_2=l_3$. This corresponds to normalized vectors $y$ that fulfill equations \eqref{eqnorms1}. To prove point 2) of the proposition, let us now consider the case $l_1l_2l_3=0$. We shall focus on the case $l_1=0$, $l_2=1$, where we have given up the normalization assumption for $y$ for a while. For other combinations of vanishing/non-vanishing $l_i$, arguments follow analogously. In the case of our interest, the matrix $\left<.\otimes y\right|C_{\Phi_t}\left|.\otimes y\right>$ equals
\begin{equation}\label{eqCPhiY0}
\left[\begin{array}{ccc}
b_t+c_tl_3&0&0\\
0&a_t+b_tl_3&-e^{i\left(\phi_3-\phi_2\right)}\sqrt{l_3}\\
0&-e^{i\left(\phi_2-\phi_3\right)}\sqrt{l_3}&c_t+a_tl_3
\end{array}\right],
\end{equation}
where $\phi_2$ and $\phi_3$ are defined as in equation \eqref{eqPartConjY}. Because the above matrix is a direct sum of a $1\times 1$ and $2\times 2$ matrix, and the individual coefficients are of degree at most $1$ in $l_3$, it is very easy to find the values of $l_3$ corresponding to singular $\left<.\otimes y\right|C_{\Phi_t}\left|.\otimes y\right>$. It turns out that the only solution when $t\in\left(0,1\right)$ is 
\begin{equation}
l_3=\frac{1}{\sqrt{t}}
\end{equation}
This corresponds to the following family of normalized vectors $y$,
\begin{equation}\label{eqY10}
\left[y^1,y^2,y^3\right]=\left[0,\sqrt{\frac{t}{1+t}}e^{i\phi_2},\sqrt{\frac{1}{1+t}}e^{i\phi_3}\right]
\end{equation}
in accordance with formula \eqref{eqnorms2}. With little additional effort, one can show that formula \eqref{eqY10} extends to the case of $t=0$. By a permutation symmetry of $C_{\Phi_t}$,
\begin{equation}\label{eqPermCPhi}
\left(C_{\Phi_t}\right)_{ij,kl}=\left(C_{\Phi_t}\right)_{\gamma\left(i\right)\gamma\left(j\right),\gamma\left(k\right)\gamma\left(l\right)}
\end{equation}
with $\gamma$ equal to the three-cycle $1\rightarrow 2\rightarrow 3\rightarrow 1$, we see that indeed formulas \eqref{eqnorms3} and \eqref{eqnorms4} describe the only remaining choices of the moduli of the coordinates of $y$ that result in singular $\left<.\otimes y\right|C_{\Phi_t}\left|.\otimes y\right>$. In other words, point 2) of the proposition holds. To prove point 3), one only needs to show that the matrices $\left<.\otimes y\right|C_{\Phi_t}\left|.\otimes y\right>$ are of rank $2$ for the three choices \eqref{eqnorms1}-\eqref{eqnorms4} of moduli of the coordinates of $y$ referred to in point 2). We leave this as an elementary algebra exercise for the reader. Points 4), 5) and 6) follow because for every $y$ such that $\mathrm{det}\left(\left<.\otimes y\right|C_{\Phi_t}\left|.\otimes y\right>\right)=0$, there exists, up to scaling, exactly one $x$ such that $\left<x\otimes y\right|C_{\Phi_t}\left|x\otimes y\right>=0$ (cf. point 3)). Moreover, these $x$'es never repeat for different $y$'s. Using block positivity of $C_{\Phi_t}$, we get
\begin{equation}
\forall_{x\in\mathbb{C}^3}!\exists_{y\in\mathbb{C}^3}\left<x\otimes y\right|C_{\Phi_t}\left|x\otimes y\right>=0
\end{equation}
which immediately proves that $\mathrm{rk}\left(\left<x\otimes .\right|C_{\Phi_t}\left|x\otimes .\right>\right)\geqslant 2$ as well as the assertions of points 4) and 3). To see that the $x$'es really never repeat, one can simply determine the form of the vectors in the kernel of $\left<.\otimes y\right|C_{\Phi_t}\left|.\otimes y\right>$ and see that they really never coincide for different $y$. For $y$ that fulfills relationships \eqref{eqnorms1}, i.e.
\begin{equation}
y=\left[e^{i\phi_1},e^{i\phi_2},e^{i\phi_3}\right]/\sqrt{3}
\end{equation}
the matrix $\left<.\otimes y\right|C_{\Phi_t}\left|.\otimes y\right>$ takes the form \eqref{eqPartConjY}. Elementary algebra shows that the kernel of this matrix consists of vectors proportional to
\begin{equation}\label{eqEx1}
x=\left[e^{-i\phi_1},e^{-i\phi_2},e^{-i\phi_3}\right]/\sqrt{3}
\end{equation}
Similarly, for $y$ that fulfills the relationships \eqref{eqnorms2}, i.e.
\begin{equation}
y=\left[0,e^{i\phi_2}\sqrt{\frac{t}{1+t}},e^{i\phi_3}\sqrt{\frac{1}{1+t}}\right]
\end{equation}
the matrix $\left<.\otimes y\right|C_{\Phi_t}\left|.\otimes y\right>$ takes the form \eqref{eqCPhiY0}. Again, elementary algebra shows that the kernel of this matrix consists of vectors proportional to 
\begin{equation}\label{eqEx2}
x=\left[0,e^{-i\phi_2}\sqrt{\frac{1}{1+t}},e^{-i\phi_3}\sqrt{\frac{t}{1+t}}\right]
\end{equation}
By the permutation symmetry \eqref{eqPermCPhi}, we also have
\begin{equation}\label{eqEx3}
x=\left[e^{-i\phi_1}\sqrt{\frac{t}{1+t}},0,e^{-i\phi_3}\sqrt{\frac{1}{1+t}}\right]
\end{equation}
corresponding to
\begin{equation}
y=\left[e^{i\phi_1}\sqrt{\frac{1}{1+t}},0,e^{i\phi_3}\sqrt{\frac{t}{1+t}}\right]
\end{equation}
and
\begin{equation}\label{eqEx4}
x=\left[e^{-i\phi_1}\sqrt{\frac{1}{1+t}},e^{-i\phi_2}\sqrt{\frac{t}{1+t}},0\right]
\end{equation}
corresponding to
\begin{equation}
y=\left[e^{i\phi_1}\sqrt{\frac{t}{1+t}},e^{i\phi_2}\sqrt{\frac{1}{1+t}},0\right]
\end{equation}
Note that the above $x$'es and $y$'s agree with formulas {(8)-(11)} in \cite{HaKye11}. It is now clear that the $x$'es given by formulas \eqref{eqEx1}, \eqref{eqEx2}, \eqref{eqEx3} and \eqref{eqEx4} can never be proportional to each other, as long as the corresponding $y$'s differ. As explained above, this implies points 4), 5) and 6) of the proposition.
\end{proof}
\end{proposition}
Due to the above proposition and Proposition \ref{propPoints}, the nonexistence of matrices $R,S\in M_3$ such that
\begin{equation}\label{eqCPhiAdABCPhi}
C_{\Phi_{t_1}}=\mathsf{Ad}_{R\otimes S}C_{\Phi_{t_2}}
\end{equation}
for arbitrary $t_1\neq t_2$, $t_1,t_2\in\left[0,1\right)$ would imply the main result of this paper, cf. Proposition \ref{propAdAB} and the discussion above equation \eqref{eqNexists}. In the following we show that indeed equality \eqref{eqCPhiAdABCPhi} cannot hold for any $R,S\in M_3$ and $t_1\neq t_2$.
\begin{proposition}\label{propLocIneq}
The Choi matrices of the Ha-Kye maps \cite{HaKye11} defined by formula \eqref{eqHaKyeFamily} with $a_t=(1-t)^2/(1-t+t^2)$, $b_t=t^2/(1-t+t^2)$, $c_t=1/(1-t+t^2)$, $t\in\left[0,1\right)$ are not locally equivalent for different $t$, i.e. equality \eqref{eqCPhiAdABCPhi} does not hold for any $R,S\in M_3$ if $t_1\neq t_2$, $t_1,t_2\in\left[0,1\right)$.
\begin{proof}
By nondecomposability of $C_{\Phi_t}$ for all $t\in\left[0,1\right)$, we get that the hypothetical matrices $R$, $S$ in \eqref{eqCPhiAdABCPhi} would have to be invertible, cf. Proposition \ref{propPoints}. If the equality really held, the following equivalence would also need to be true
\begin{equation}
F_{t_1}\left(y\right)=0\Leftrightarrow F_{t_2}\left(S^{\ast}y\right)=0
\end{equation}
cf. the definition of $F_t$, eq. \eqref{eqFt}. Let us consider $y=\left[e^{i\phi_1},e^{i\phi_2},e^{i\phi_3}\right]/\sqrt{3}$ for arbitrary $\phi_1$, $\phi_2$ and $\phi_3$, which is an admissible choice of $y$ according to the proof of Proposition \ref{propPoints2}. We now show that none of the moduli of the coordinates of $S^{\ast}y$ can vanish for any choice of the phases $\phi_j$, $j=1,2,3$. Indeed, according to the proof of Proposition \ref{propPoints2}, at most two of them could possibly vanish for some choice of $\phi_j$'s to ensure $\mathrm{det}\left(\left<.\otimes S^{\ast}y\right|C_{\Phi_{t_2}}\left|.\otimes S^{\ast}y\right>\right)=0$. Let the non-vanishing coordinate be the third one and assume that the first one vanishes. Since $S^{\ast}$ is non-singular, by changing one of the $\phi_j$'s by a sufficiently small $\delta$, we can make the first coordinate non-negative without changing the third coordinate significantly. Hence, the ratio of the modulus of the first coordinate to the modulus of the second coordinate would have to become different from zero. Thus, by continuity of the moduli of the coordinates of $S^{\ast}y$ as functions of $\phi_j$'s, the ratio of the first coordinate to the second coordinate would need to take an infinite number of values as a function of the $\delta$. However, this is not possible if $\mathrm{det}\left(\left<.\otimes S^{\ast}y\right|C_{\Phi_{t_2}}\left|.\otimes S^{\ast}y\right>\right)$ is assumed to vanish for all $\phi_j$'s, since by the proof of Proposition \ref{propPoints2}, only three values of the ratio are allowed, namely $0$, $\sqrt{t_2}$ and $1$, cf. equations \eqref{eqEx1}-\eqref{eqEx4}. In this way we have shown that the moduli of the coordinates of $S^{\ast}y$ for $y$ of the form $\left[e^{i\phi_1},e^{i\phi_2},e^{i\phi_3}\right]/\sqrt{3}$ must all be non-vanishing, no matter what the choice of the $\phi_j$'s is. Let us note that these coordinates equal
\begin{equation}
\frac{1}{\sqrt{3}}\sum_{l=1}^3\overline{S_{lk}}e^{i\phi_l}
\end{equation}
for $k=1,2,3$. The numbers $S_{lk}$ are the matrix coefficients of $S$. According to an argument presented in the proof of Proposition \ref{propPoints2}, cf. equations \eqref{eqEx1}-\eqref{eqEx4}, the only possibility for $\left<.\otimes S^{\ast}y\right|C_{\Phi_{t_2}}\left|.\otimes S^{\ast}y\right>$ to be singular for all choices of the $\phi_j$'s is when
\begin{equation}\label{eqEqModuluses}
\left|\sum_{l=1}^3\overline{S_{l1}}e^{i\phi_l}\right|=\left|\sum_{l=1}^3\overline{S_{l2}}e^{i\phi_l}\right|=\left|\sum_{l=1}^3\overline{S_{l3}}e^{i\phi_l}\right|
\end{equation} 
for all choices of $\left\{\phi_j\right\}_{j=1,2,3}$. With little additional effort, one can show that the above equalities can can only hold if the matrix $S^{\ast}$ has exactly one nonzero entry in each row or that some of its rows must be proportional to each other (cf. Appendix \ref{appModuluses}). Because in our setting $S^{\ast}$ is necessarily non-singular, it must therefore be a permutation matrix multiplied by a non-singular diagonal matrix. We now show that the permutation part is necessarily a three-cycle or the identity permutation. To exclude transpositions as the permutation part, it is enough to observe that the specific vectors $y$ of the form \eqref{eqEx2}, \eqref{eqEx3} and \eqref{eqEx4} with $t=t_1$ cannot all transform to some of their legitimate counterparts for $t=t_2$, i.e. some vectors of the form \eqref{eqEx2}, \eqref{eqEx3} and \eqref{eqEx4}, or possibly \eqref{eqEx1}, with $t=t_2$. We provide a detailed argument for the transposition $2\leftrightarrow 3$ and skip the other ones because they follow by trivial analogy. Let us therefore assume that 
\begin{equation}\label{eqSastZT}
S^{\ast}=\left[\begin{array}{ccc}
\zeta_1&0&0\\
0&\zeta_2&0\\
0&0&\zeta_3
\end{array}
\right]\left[\begin{array}{ccc}
1&0&0\\
0&0&1\\
0&1&0
\end{array}
\right]
\end{equation}
where $\zeta_1,\zeta_2,\zeta_3\in\mathbb{Z}$, $\zeta_1\zeta_2\zeta_3\neq 0$.
For $y$ of the form \eqref{eqEx2} with $t=t_1$, we have
\begin{equation}
S^{\ast}y=\left[0,\zeta_2e^{i\phi_2}\sqrt{\frac{1}{1+t_1}},\zeta_3e^{i\phi_3}\sqrt{\frac{t_1}{1+t_1}}\right]
\end{equation}
According to Proposition \ref{propPoints2}, the only possibility for $\left<.\otimes S^{\ast}y\right|C_{\Phi_2}\left|.\otimes S^{\ast}y\right>$ to be singular is when
\begin{eqnarray}
\zeta_2e^{i\phi_2}\sqrt{\frac{1}{1+t_1}}&=&\alpha e^{i\phi'_2}\sqrt{\frac{t_2}{1+t_2}}\label{eqFracFrac1}\\
\zeta_3e^{i\phi_3}\sqrt{\frac{t_1}{1+t_1}}&=&\alpha e^{i\phi'_3}\sqrt{\frac{1}{1+t_2}}\label{eqFracFrac2}
\end{eqnarray}
for some phases $\phi'_2$, $\phi'_3$ and a scaling factor $\alpha\in\mathbb{R}\setminus\left\{0\right\}$. In the above formula, we used the fact that at least one of the parameters $t_1$, $t_2$ is nonzero.
Similarly, by considering the transforms $S^{\ast}y$ for $y$ of the form \eqref{eqEx3} and \eqref{eqEx4}, we arrive at the following equations for $t_1$, $t_2$ and the $\zeta_i$'s
\begin{eqnarray}  
\zeta_1e^{i\phi_1}\sqrt{\frac{1}{1+t_1}}&=&\beta e^{i\phi''_1}\sqrt{\frac{t_2}{1+t_2}}\label{eqFracFrac3}\\
\zeta_2e^{i\phi_3}\sqrt{\frac{t_1}{1+t_1}}&=&\beta e^{i\phi''_2}\sqrt{\frac{1}{1+t_2}}\label{eqFracFrac4}
\end{eqnarray}
and
\begin{eqnarray}
\zeta_1e^{i\phi_1}\sqrt{\frac{t_1}{1+t_1}}&=&\gamma e^{i\phi_1'''}\sqrt{\frac{1}{1+t_2}}\label{eqFracFrac5}\\
\zeta_3e^{i\phi_2}\sqrt{\frac{1}{1+t_1}}&=&\gamma e^{i\phi_3'''}\sqrt{\frac{t_2}{1+t_2}}\label{eqFracFrac6}
\end{eqnarray}
where $\phi''_1$, $\phi''_2$, $\phi'''_1$ and $\phi'''_3$ are again some arbitrary phases and $\beta,\gamma\in\mathbb{R}\setminus\left\{0\right\}$ - arbitrary scaling factors.

By calculating the moduli of \eqref{eqFracFrac2} and \eqref{eqFracFrac6}, we get the following conditions on $t_1\in\left(0,1\right)$, $t_2\in\left(0,1\right)$ and $\zeta_3\neq 0$
\begin{eqnarray}
\left|\zeta_3\right|\sqrt{\frac{1}{1+t_1}}&=&\left|\gamma\right|\sqrt{\frac{t_2}{1+t_2}}\\
\left|\zeta_3\right|\sqrt{\frac{t_1}{1+t_1}}&=&\left|\alpha\right|\sqrt{\frac{1}{1+t_2}}
\end{eqnarray}
Clearly, the above formulas can only hold if
\begin{equation}\label{eqtt1}
t_1t_2=\left|\frac{\alpha}{\gamma}\right|^2
\end{equation}
In a similar way, one can obtain
\begin{equation}\label{eqtt2}
t_1t_2=\left|\frac{\beta}{\alpha}\right|^2
\end{equation} 
and
\begin{equation}\label{eqtt3}
t_1t_2=\left|\frac{\gamma}{\beta}\right|^2
\end{equation} 
The above formulas imply
\begin{equation}
t_1t_2=1
\end{equation}
which is not true for any $t_1,t_2\in\left(0,1\right)$, $t_1\neq t_2$. Thus matrices $S^{\ast}$ of the form \eqref{eqSastZT} are not good candidates for the transform $S^{\ast}$. By trivial analogy, the same conclusion holds when the particular transposition matrix in \eqref{eqSastZT} is replaced with another one. Thus we have proved that the only possible form of the matrix $S^{\ast}$ is  
\begin{equation}\label{eqSastZC1}
S^{\ast}=\left[\begin{array}{ccc}
\zeta_1&0&0\\
0&\zeta_2&0\\
0&0&\zeta_3
\end{array}
\right]\left[\begin{array}{ccc}
0&0&1\\
1&0&0\\
0&1&0
\end{array}
\right]
\end{equation}
or 
\begin{equation}\label{eqSastZC2}
S^{\ast}=\left[\begin{array}{ccc}
\zeta_1&0&0\\
0&\zeta_2&0\\
0&0&\zeta_3
\end{array}
\right]\left[\begin{array}{ccc}
0&1&0\\
0&0&1\\
1&0&0
\end{array}
\right]
\end{equation}
or 
\begin{equation}\label{eqSastZC3}
S^{\ast}=\left[\begin{array}{ccc}
\zeta_1&0&0\\
0&\zeta_2&0\\
0&0&\zeta_3
\end{array}
\right]
\end{equation}
It turns out that for matrices $S^{\ast}$ of the form \eqref{eqSastZC1}, \eqref{eqSastZC2} and \eqref{eqSastZC3}, the equality \eqref{eqCPhiAdABCPhi} can only hold if $t_1=t_2$. In the following, we prove this for $S^{\ast}$ of the form \eqref{eqSastZC1}. For the cases of \eqref{eqSastZC2} and \eqref{eqSastZC3}, the reasoning follows analogously and we leave the details to the interested reader. Once we have convinced ourselves that matrices of the form \eqref{eqSastZC1}, \eqref{eqSastZC2} and \eqref{eqSastZC3} are of no use in transforming different members of the family $\left\{\Phi_t\right\}_{t\in\left[0,1\right)}$ between each other, we will have completed the proof of the proposition. 

Let us assume that $S^{\ast}$ is of the form \eqref{eqSastZC1}. We again consider the transforms $S^{\ast}y$ for the characteristic vectors $y$ of the form \eqref{eqEx2}, \eqref{eqEx3} and \eqref{eqEx4} with $t=t_1$. Just as before, these must be again of the form \eqref{eqEx2}, \eqref{eqEx3}, \eqref{eqEx4}, or possibly \eqref{eqEx1}, for $t=t_2$. If $y$ is of the form \eqref{eqEx2}, the transform $S^{\ast}$
equals
\begin{equation}\label{eqSastEx2}
S^{\ast}y=\left[\zeta_1e^{i\phi_3}\sqrt{\frac{1}{1+t_1}},0,\zeta_3e^{i\phi_2}\sqrt{\frac{t_1}{1+t_1}}\right]
\end{equation}
Under the assumption of $t_1\neq 0$ or $t_2\neq 0$, which is true when $t_1\neq t_2$, the requirement that $\mathrm{det}\left(\left<.\otimes S^{\ast}y\right|C_{\Phi_2}\left|.\otimes S^{\ast}y\right>\right)$ vanishes can only be met if the $S^{\ast}y$ given by \eqref{eqSastEx2} coincides with
\begin{equation}
\alpha\left[e^{i\phi'_1}\sqrt{\frac{1}{1+t_2}},0,e^{i\phi'_3}\sqrt{\frac{t_2}{1+t_2}}\right]
\end{equation}
for some phases $\phi'_1$ and $\phi'_3$ and a scaling factor $\alpha\in\mathbb{R}\setminus\left\{0\right\}$. This clearly implies
\begin{eqnarray}
\zeta_1e^{i\phi_3}\sqrt{\frac{1}{1+t_1}}=\alpha e^{i\phi'_1}\sqrt{\frac{1}{1+t_2}}\label{eqFracFrac7}\\
\zeta_3e^{i\phi_2}\sqrt{\frac{t_1}{1+t_1}}=\alpha e^{i\phi'_3}\sqrt{\frac{t_2}{1+t_2}}\label{eqFracFrac8}
\end{eqnarray}
Similarly, by considering the transforms $S^{\ast}y$ for vectors $y$ of the form \eqref{eqEx3} and \eqref{eqEx4}, we arrive at the conditions
\begin{eqnarray}
\zeta_1e^{i\phi_3}\sqrt{\frac{t_1}{1+t_1}}=\beta e^{i\phi''_1}\sqrt{\frac{t_2}{1+t_2}}\label{eqFracFrac9}\\
\zeta_2e^{i\phi_1}\sqrt{\frac{1}{1+t_1}}=\beta e^{i\phi''_2}\sqrt{\frac{1}{1+t_2}}\label{eqFracFrac10}
\end{eqnarray}
and
\begin{eqnarray}
\zeta_2e^{i\phi_1}\sqrt{\frac{t_1}{1+t_1}}=\gamma e^{i\phi'''_2}\sqrt{\frac{t_2}{1+t_2}}\label{eqFracFrac11}\\
\zeta_3e^{i\phi_2}\sqrt{\frac{1}{1+t_1}}=\gamma e^{i\phi'''_3}\sqrt{\frac{1}{1+t_2}}\label{eqFracFrac12}
\end{eqnarray}
where $\phi''_1$, $\phi''_2$, $\phi'''_2$ and $\phi'''_3$ denote some arbitrary phases and $\beta,\gamma\in\mathbb{R}\setminus\left\{0\right\}$ are arbitrary nonzero scaling factors. By considering the moduli of \eqref{eqFracFrac7} and \eqref{eqFracFrac8}, we easily obtain
\begin{equation}
\left|\zeta_1\right|^2t_2=\left|\zeta_3\right|^2t_1\label{eqt1t2_1}
\end{equation}
In a similar way, from equations \eqref{eqFracFrac9}-\eqref{eqFracFrac12} we can get
\begin{eqnarray}
\left|\zeta_2\right|^2t_2&=&\left|\zeta_1\right|^2t_1\label{eqt1t2_2}\\
\left|\zeta_3\right|^2t_2&=&\left|\zeta_2\right|^2t_1\label{eqt1t2_3}
\end{eqnarray}
We can now multiply equalities \eqref{eqt1t2_1}-\eqref{eqt1t2_3} together and get
\begin{equation}
\left|\zeta_1\right|^2\left|\zeta_2\right|^2\left|\zeta_3\right|^2t_2^3=\left|\zeta_1\right|^2\left|\zeta_2\right|^2\left|\zeta_3\right|^2t_1^3\Rightarrow t_2=t_1
\end{equation} 
This clearly contradicts $t_1\neq t_2$ and proves the main assertion of the proposition.
\end{proof} 
\end{proposition}
As we already explained above, the above proposition implies the main result of this paper. We briefly remind the reader why this is the case. By Proposition \ref{propPoints2}, the family $\left\{\Phi_t\right\}_{t\in\left[0,1\right)}$ of extreme indecomposable maps of the form \eqref{eqHaKyeFamily}, as considered by Ha and Kye \cite{HaKye11}, fulfills the assumptions of point 5) of Proposition \ref{propPoints}. Therefore, if equality \eqref{eqConvhull} holds for some positive maps $\left\{\Xi_i\right\}_{i=1}^k\subset\mathcal{P}_{3,3}$, there must exist at least two values $t_1,t_2\in\left[0,1\right)$, $t_1\neq t_2$ such that 
\begin{equation}\label{eqAdABPhiPhi}
\Phi_{t_1}=\mathsf{Ad}_A\circ\Phi_{t_2}\circ\mathsf{Ad}_B
\end{equation}
for some non-singular $A,B\in M_3$ (actually, there must exist an infinite number of such pairs). By Proposition \ref{propAdAB}, we know that \eqref{eqAdABPhiPhi} is equivalent to \eqref{eqCPhiAdABCPhi} with $R=B^t$, $S=A$. However, we know from Proposition \ref{propLocIneq} that \eqref{eqCPhiAdABCPhi} does not hold for any $t_1\neq t_2$, $R,S\in M_3$. Thus equality \eqref{eqConvhull} cannot be true. By Proposition \ref{propEquivConvhull}, this is equivalent to the non-existence of a necessary and sufficient separability criterion of the form \eqref{eqCritFamily3}. In this way we have arrived at the following conclusion
\begin{theorem}\label{thmMain}
There does not exist a separability criterion of the form
\begin{equation}\label{eqCritFamilyX}
\rho\in\mathsf{Sep}_{3\times 3}\Leftrightarrow\forall_{i=1,2,\ldots,k}\forall_{B\in M_3}\left(\mathbb{I}\otimes\Phi_i\right)\mathsf{Ad}_{\mathbb{I}\otimes B}\rho\geqslant 0
\end{equation}
for a finite set of positive maps $\left\{\Phi_i\right\}_{i=1,2,\ldots,k}$.
\end{theorem}
For a more mathematically oriented audience, the same can be expressed, by Proposition \ref{propEquivConvhull}, as
\begin{theorem}\label{thmMain2}
The cone $\mathcal{P}_{3,3}$ of positive maps from the set $M_3$ of three-dimensional matrices into itself is not finitely generated as a mapping cone. In other words, there does not exist a finite set of positive maps $\left\{\Xi_i\right\}_{i=1,2,\ldots,k}$ such that 
\begin{equation}\label{eqConvhullX}
\mathcal{P}_{3,3}=\overline{\mathrm{convhull}\left(\left\{\mathsf{Ad}_X\circ\Xi_i\circ\mathsf{Ad}_Y|i;X,Y\in M_3\right\}\right)},
\end{equation}
holds. 
\end{theorem}
Some other noteworthy properties of mapping cones can be found in \cite{Stormer08,Stormer09,Stormer09II,SSZ09,Skowronek11,JohnstonStormer12,JSS12}.

\section{Conclusion}
It is a generally accepted opinion that the detection of all density matrix entanglement in $3\times 3$ and more complex systems is a difficult task. In this paper we have shown that for such systems there is no hope to have a sufficient and necessary criterion for separability of a state, similar in structure to the famous positive partial transpose for $2\times 2$ and $2\times 3$ systems \cite{Peres96,Horodeccy96}. This is even true despite the fact that we allow arbitrary local transforms in formula \eqref{eqCritFamily3}. The argument we presented also suggests that in order to have a strong entanglement detection procedure, one should at least consider a criterion of the form \eqref{eqCritFamily3} with $\left\{\Xi_i\right\}_{i=1}^k$ substituted with the whole Ha-Kye family of positive maps $\left\{\Phi_t\right\}_{t\in\left[0,1\right)}$ \cite{HaKye11}.

From a more mathematics centered point of view, our result shows that the cone $\mathcal{P}_{3,3}$ of positive maps from the set of three-dimensional matrices into itself, is not finitely generated as a mapping cone. This is in contrast with $\mathcal{P}_{3,2}$, $\mathcal{P}_{2,3}$ and $\mathcal{P}_{2,2}$, where all positive maps turn out to be decomposable \cite{Stormer63,Woronowicz76}.    

\begin{remark}
It is natural to expect that the cones $\mathcal{P}_{m,n}$ of positive maps with $m\geqslant 3$ and $n\geqslant 3$ are also not finitely generated as mapping cones, however the author has not yet found a proof of this fact. If it holds, we immediately obtain a generalization of the main result of this paper to $m\times n$ systems where $m,n\geqslant 3$.
\end{remark}

\begin{remark}
In the light of the proof presented above, it is natural to ask whether the whole family of Ha-Kye maps can provide us with a necessary and sufficient separability criterion of the form \eqref{eqCritFamily3}. 
\end{remark}

\section*{Acknowledgement}
The result presented in this paper is a byproduct of work done together with Erling St{\o}rmer. The author is heartily indebted to Karol Życzkowski, whose emails finally convinced him write this paper up, even after a few years spent outside academia. Valuable comments by Martyna Rymer are gratefully acknowledged. This work has been financially supported by the NCN Grant no. 2015/18/A/ST2/00274. 

\appendix

\section{Choi matrix change under multiplication by conjugation maps}\label{appAdAB}

We would like to show that
\begin{equation}
C_{\mathsf{Ad}_A\circ\Phi\circ\mathsf{Ad}_B}=\mathsf{Ad}_{B^t\otimes A}C_{\Phi}
\end{equation}
for arbitrary $A,B\in M_3$ and a positive map $\mathcal{P}_{3,3}$. As expected, this result holds for general $\mathcal{P}_{m,n}$ as well.

From the definition \eqref{eqChoi} of the Choi matrix, we immediately see that
\begin{equation}
C_{\mathsf{Ad}_A\circ\Phi\circ\mathsf{Ad}_B}=\mathsf{Ad}_{\mathbb{I}\otimes A}C_{\Phi\circ\mathsf{Ad}_B}
\end{equation}
Thus, we only need to prove that 
\begin{equation}
C_{\Phi\circ\mathsf{Ad}_B}=\mathsf{Ad}_{B^t\otimes\mathbb{I}}C_{\Phi}
\end{equation}
We have
\begin{equation}
C_{\Phi\circ\textsf{Ad}_B}=\left(\mathbb{I}\otimes\Phi\right)\sum_{i,j}\sum_{k,l}e_{ij}\otimes B_{ki}\overline{B_{lj}}e_{kl},
\end{equation}
where $B_{ij}$ are the matrix coefficients of $B$. We further have
\begin{equation}
\sum_{i,j}\sum_{k,l}e_{ij}\otimes B_{ki}\overline{B_{lj}}e_{kl}=\sum_{i,j}\sum_{k,l}B_{ki}\overline{B_{lj}}e_{ij}\otimes e_{kl},
\end{equation}
which is the same as
\begin{equation}
\mathsf{Ad}_{B^t\otimes\mathbb{I}}\sum_{k,l}e_{kl}\otimes e_{kl}
\end{equation}
Thus we can write
\begin{equation}
C_{\Phi\circ\mathsf{Ad}_B}=\left(\mathbb{I}\otimes\Phi\right)\mathsf{Ad}_{B^t\otimes\mathbb{I}}\sum_{k,l}e_{kl}\otimes e_{kl}
\end{equation}
From this, we easily obtain
\begin{equation}
C_{\Phi\circ\mathsf{Ad}_B}=\mathsf{Ad}_{B^t\otimes\mathbb{I}}\sum_{k,l}e_{kl}\otimes\Phi\left(e_{kl}\right)=\mathsf{Ad}_{B^t\otimes\mathbb{I}}C_{\Phi}
\end{equation}

\section{Extreme positive maps in a finitely generated mapping cone}\label{appClosed}
In this appendix we prove that extreme positive maps $\Phi\in\mathcal{P}_{3,3}$ in a finitely generated closed mapping cone of the form \eqref{eqConvhull} must either be of the form $\mathsf{Ad}_X\circ\Xi_i\circ\mathsf{Ad}_Y$ for some $i\in\left\{1,2,\ldots,k\right\}$ and $X,Y\in M_3$ or there must exist $x\in\mathbb{C}^3$ or $y\in\mathbb{C}^3$ such that
\begin{equation}\label{eqRk1}
\mathsf{rk}\left(\left<x\otimes .\right|C_{\Phi}\left|x\otimes .\right>\right)<2
\end{equation}
or
\begin{equation}\label{eqRk2}
\mathsf{rk}\left(\left<.\otimes y\right|C_{\Phi}\left|.\otimes y\right>\right)<2\\
\end{equation}
Point 3) of Proposition \ref{propPoints} trivially follows from this result.

Let us assume that $\Phi$ is an element of the closed convex cone \eqref{eqConvhull}. Equivalently, there exist sequences $\left\{i^l_j\right\}_{l=1}^{+\infty}\subset\mathbb{N}$, $\left\{X^l_j\right\}_{l=1}^{+\infty}\subset M_3$ and $\left\{Y^l_j\right\}_{l=1}^{+\infty}\subset M_3$, for $j=1,2,\ldots,82$ such that 
\begin{equation}\label{eqAdLimit}
\sum_{j=1}^{82}\mathsf{Ad}_{X^l_j}\circ\Xi_{i^l_j}\circ\mathsf{Ad}_{Y^l_j}\xrightarrow{l\rightarrow +\infty}\Phi
\end{equation}
The possibility to set an upper border of $82$ on summation by $j$ follows from Carath\'eodory's theorem \cite{Rockafellar}.

Note that every non-zero positive map $\Phi$ satisfies $\mathrm{Tr}\left(C_{\Phi}\right)>0$ and hence can be normalized to have trace of its Choi matrix equal to one. Moreover, positive maps $\Phi$ such that $C_{\Phi}$ have unit trace, form a compact subset of the subspace of all maps $\Psi$ in $\mathcal{H}_{m,n}$ with the trace of $C_{\Psi}$ equal to one. In that subspace, equation \eqref{eqAdLimit} induces
\begin{equation}\label{eqAdLimitNorm}
\sum_{j=1}^{82}\lambda^l_j\frac{\mathsf{Ad}_{X^l_j}\circ\Xi_{i^l_j}\circ\mathsf{Ad}_{Y^l_j}}{\tau^l_j}\xrightarrow{l\rightarrow +\infty}\frac{\Phi}{\mathrm{Tr}\left(C_{\Phi}\right)},
\end{equation} 
where $\tau^l_j:=\mathrm{Tr}\left(C_{\mathsf{Ad}_{X^l_j}\circ\Xi_{i^l_j}\circ\mathsf{Ad}_{Y^l_j}}\right)$, $\lambda^l_j\geqslant 0$, the sum $\sum_{j=1}^{82}\lambda^l_j$ tends to $1$ for $l\rightarrow +\infty$, and we are free to assume that $\tau^l_j>0$ for all $j$ and $l$. The last assumption is permissible because by setting $X^l_j$ and $Y^l_j$ equal to $\mathbb{I}$ if initially $\mathrm{Tr}\left(C_{\mathsf{Ad}_{X^l_j}\circ\Xi_{i^l_j}\circ\mathsf{Ad}_{Y^l_j}}\right)=0$ and bringing the respective $\lambda^l_j$ to zero, we can make all the $\tau^l_j$'s positive, at the same time keeping the relationship \eqref{eqAdLimitNorm} a consequence of \eqref{eqAdLimit}. Moreover, we can assume that the largest singular values of $X^l_j$'s and $Y^l_j$'s are equal to one because it is always possible to rescale the $X^l_j$'s and $Y^l_j$'s and their singular values with no influence on $\left(\mathsf{Ad}_{X^l_j}\circ\Xi_{i^l_j}\circ\mathsf{Ad}_{Y^l_j}\right)/{\tau^l_i}$. 

In summary, equation \eqref{eqAdLimit} implies \eqref{eqAdLimitNorm} with $\lambda^l_j$'s and $\left(\mathsf{Ad}_{X^l_j}\circ\Xi_{i^l_j}\circ\mathsf{Ad}_{Y^l_j}\right)/{\tau^l_i}$'s contained in compact subsets of $\mathbb{R}$ and $\mathcal{P}_{3,3}$. Therefore, there must exist limits
\begin{eqnarray}
\frac{\mathsf{Ad}_{X^{l_w}_j}\circ\Xi_{i^{l_w}_j}\circ\mathsf{Ad}_{Y^{l_w}_j}}{\tau^l_i}&\xrightarrow{w\rightarrow +\infty}&\Sigma_j\in\mathcal{P}_{3,3}\\
\lambda^{l_w}_j&\xrightarrow{w\rightarrow +\infty}&\lambda_j\geqslant 0
\end{eqnarray}
for certain subsequence $\left\{l_w\right\}_{w=1}^{+\infty}\subset\mathbb{N}$. Hence,
\begin{equation}
\frac{\Phi}{\mathrm{Tr}\left(C_{\Phi}\right)}=\sum_{j=1}^{82}\lambda_j\Sigma_j
\end{equation}
Obviously, for $\Phi$ extreme, we can assume that the above sum only contains a single nonzero term $\Sigma$, where 
\begin{equation}\label{eqLimAd}
\Sigma=\lim\limits_{w \to +\infty}\frac{\mathsf{Ad}_{X^w}\circ\Xi_{i^w}\circ\mathsf{Ad}_{Y^w}}{\mathrm{Tr}\left(C_{\mathsf{Ad}_{X^w}\circ\Xi_{i^w}\circ\mathsf{Ad}_{Y^w}}\right)}
\end{equation}
for appropriately chosen $X^w$ and $Y^w$ with largest singular values equal to one. By an appropriate choice of a subsequence, we can also assure that $i_w$ is a constant $i\in\left\{1,2,\ldots, k\right\}$. In the following, we shall simply assume that $i_w=i$ in equation \eqref{eqLimAd}. 

We now prove that the limit on the right-hand side of equation \eqref{eqLimAd} must either be of the form $\mathsf{Ad}_X\circ\Xi_i\circ\mathsf{Ad}_Y$ for some $X,Y\in M_3$ or equation \eqref{eqRk1} or \eqref{eqRk2} with $\Sigma$ substituted for $\Phi$ must hold for some $x$ or $y$ in $\mathbb{C}^3$.

Because $X^w$ and $Y^w$ can be assumed to belong to a compact set of matrices with largest singular value equal to one, there must exist a subsequence $\left\{w_t\right\}_{t=1}^{+\infty}$ and matrices $X$, $Y$ with largest singular values equal to one, such that
\begin{eqnarray}
X^{w_t}&\xrightarrow{t\rightarrow +\infty}&X\\
Y^{w_t}&\xrightarrow{t\rightarrow +\infty}&Y
\end{eqnarray}
Without loss of generality, we may assume that $X^w$ and $Y^w$ themselves converge to $X$ and $Y$, respectively.

If $\lim\limits_{w\rightarrow +\infty}\mathrm{Tr}\left(C_{\mathsf{Ad}_{X^w}\circ\Xi_{i}\circ\mathsf{Ad}_{Y^w}}\right)=\tau>0$, our main assertion is obviously true, because we obtain
\begin{equation}
\Sigma = \mathsf{Ad}_{X/\sqrt[4]{\tau}}\circ\Xi_i\circ\mathsf{Ad}_{Y/\sqrt[4]{\tau}}
\end{equation}
and
\begin{equation}
\Phi=\mathsf{Ad}_{\sqrt[4]{\frac{\sigma}{\tau}}X}\circ\Xi_i\circ\mathsf{Ad}_{\sqrt[4]{\frac{\sigma}{\tau}}Y},
\end{equation}
where $\sigma:=\mathrm{Tr}\left(C_{\Phi}\right)>0$. Therefore, we only need to consider the case of $\mathrm{Tr}\left(C_{\mathsf{Ad}_{X^w}\circ\Xi_{i}\circ\mathsf{Ad}_{Y^w}}\right)$ that tends to zero as $w\rightarrow +\infty$.

By Proposition \ref{propAdAB}
\begin{equation}
C_{\mathsf{Ad}_{X^w}\circ\Xi_i\circ\mathsf{Ad}_{Y^w}}=\mathsf{Ad}_{\left(Y^w\right)^t\otimes X^w}C_{\Xi_i}
\end{equation}
Denote the second and the third singular value of $\left(Y^w\right)^t$ by $\sigma^w_2$, $\sigma^w_3$. Similarly, let the second and the third singular value of $X^w$ be $\delta^w_2$ and $\delta^w_3$. For convenience, we also define $\sigma^w_1=\delta^w_1:=1$. Let
\begin{eqnarray}
\left(Y^w\right)^t&=&\sum_{i=1}^3\sigma^w_i\left|u^w_i\right>\left<v^w_i\right|\label{eqSing1}\\
X^w&=&\sum_{i=1}^3\delta^w_i\left|p^w_i\right>\left<q^w_i\right|\label{eqSing2}
\end{eqnarray}
be decompositions following directly from singular value decompositions of $\left(Y^w\right)^t$ and $X^w$. In particular, $\left\{u^w_i\right\}_{i=1}^3$, $\left\{v^w_i\right\}_{i=1}^3$, $\left\{p^w_i\right\}_{i=1}^3$, $\left\{q^w_i\right\}_{i=1}^3$ in equations \eqref{eqSing1} and \eqref{eqSing2} are orthonormal bases of $\mathbb{C}^3$ with elements corresponding to rows or columns of unitary matrices present in the respective singular value decompositions.  We then have
\begin{multline}\label{eqSingSum}
C_{\mathsf{Ad}_{X^w}\circ\Xi_i\circ\mathsf{Ad}_{Y^w}}=\\=\sum_{r,s,t,u=1}^3\sigma^w_r\sigma^w_t\delta^w_s\delta^w_u\left<v^w_r\otimes q^w_s\right|C_{\Xi_i}\left|v^w_t\otimes q^w_u\right>\cdot\\\cdot\left|u^w_r\otimes p^w_s\right>\left<u^w_t\otimes p^w_u\right|
\end{multline}
With no loss of generality, we may assume that all the sequences $\left\{\sigma^w_i\right\}_{w=1}^{+\infty}$, $\left\{\delta^w_i\right\}_{w=1}^{+\infty}$, $\left\{u^w_i\right\}_{w=1}^{+\infty}$, $\left\{v^w_i\right\}_{w=1}^{+\infty}$, $\left\{p^w_i\right\}_{w=1}^{+\infty}$, $\left\{q^w_i\right\}_{w=1}^{+\infty}$ are convergent. This is so because there always exists their common convergent subsequence, which we can keep and remove the other terms. It is then easy to see that in an appropriate orthonormal basis, the $\left(r,s,t,u\right)$-th matrix element of $C_{\mathsf{Ad}_{X^w}\circ\Xi_i\circ\mathsf{Ad}_{Y^w}}/\mathrm{Tr}\left(C_{\mathsf{Ad}_{X^w}\circ\Xi_i\circ\mathsf{Ad}_{Y^w}}\right)$ equals
\begin{equation}\label{eqLimFrac}
\lim\limits_{w \to +\infty}\frac{\sigma^w_r\sigma^w_t\delta^w_s\delta^w_u\left<v^w_r\otimes q^w_s\right|C_{\Xi_i}\left|v^w_t\otimes q^w_u\right>}{\sum_{h,j=1}^3\left(\sigma^w_h\right)^2\left(\delta^w_j\right)^2\left<v^w_h\otimes q^w_j\right|C_{\Xi_i}\left|v^w_h\otimes q^w_j\right>}
\end{equation}
Note also that we are free to assume the existence of a subsequence $\left\{w_n\right\}_{n=1}^{+\infty}$ such that $\sigma^{w_n}_r,\sigma^{w_n}_t,\delta^{w_n}_s,\delta^{w_n}_u>0$. Otherwise, we could choose one of the singular values to be equal to $0$ all the time on an appropriate subsubsequence and the existence of $x$ or $y$ such that equations \eqref{eqRk1} or \eqref{eqRk2} hold would immediately follow (actually, $C_{\Phi}$ would turn out to be supported on a $2\times 3$ or smaller subspace). As a consequence of that, with no loss of generality, we may assume $\sigma^{w}_r,\sigma^{w}_t,\delta^{w}_s,\delta^{w}_u>0$.

The fact that the denominator of \eqref{eqLimFrac} vanishes in the limit $w\rightarrow +\infty$, together with $\sigma_1=\delta_1=1$ implies that $\left<v_1\otimes q_1\right|C_{\Xi_i}\left|v_1\otimes q_1\right>=0$. Let us now focus on the terms with $\sigma^w_2$, although the same argument goes for any other of the singular values $\sigma^w_2$, $\sigma^w_3$, $\delta^w_2$ and $\delta^w_3$. If one of the corresponding matrix elements of $\Xi_i$, $\left<v_2\otimes q_1\right|C_{\Xi_i}\left|v_2\otimes q_1\right>$, equals $0$, the following lemma shows that the main assertion of this appendix is true because \eqref{eqRk2} holds for some $y$.
\begin{lemma}
Let $\left<v_1\otimes q_j\right|C_{\Xi_i}\left|v_1\otimes q_j\right>=0$ and $\left<v_2\otimes q_j\right|C_{\Xi_i}\left|v_2\otimes q_j\right>=0$ for some $j\in\left\{1,2,3\right\}$. Then there exists $y\in\mathbb{C}^3$ such that
\begin{equation}\label{eqRkyCPhi}
\mathrm{rk}\left(\left<.\otimes y\right|C_{\Phi}\left|.\otimes y\right>\right)=\mathrm{rk}\left(\left<.\otimes y\right|C_{\Sigma}\left|.\otimes y\right>\right)<2
\end{equation}
\begin{proof}
Let us decompose the vector $q_j$ in the basis $\left\{q^w_u\right\}_{u=1}^{3}$,
\begin{equation}
q_j=\sum_{u=1}^3\lambda^w_{ju}q^w_u
\end{equation}
Now define
\begin{equation}
y^w:=\frac{\sum_{u=1}^3\frac{\lambda^w_{ju}}{\delta^w_u}p^w_u}{\sqrt{\sum_{u=1}^3\left|\frac{\lambda^w_{ju}}{\delta^w_u}\right|^2}}
\end{equation}
From equation \eqref{eqSingSum}, we easily see that
\begin{equation}\label{eqRkyw}
\mathrm{rk}\left(\left<.\otimes y^w\right|\frac{C_{\mathsf{Ad}_{X^w}\circ\Xi_i\circ\mathsf{Ad}_{Y^w}}}{\mathrm{Tr}\left(C_{\mathsf{Ad}_{X^w}\circ\Xi_i\circ\mathsf{Ad}_{Y^w}}\right)}\left|.\otimes y^w\right>\right)<2,
\end{equation}
where we have used the fact that for the vanishing of $\left<x_1\otimes y^w\right|X\left|x_1\otimes y^w\right>$ and $\left<x_2\otimes y^w\right|X\left|x_2\otimes y^w\right>$ for a block positive matrix $X$ implies $\mathrm{rk}\left(\left<.\otimes y^w\right|X\left|.\otimes y^w\right>\right)<2$, as long as $x_1\neq x_2$.

Since $y^w$'s belong to a compact subset of $\mathbb{C}^3$ consisting of normalized vectors, there is a subsequence $\left\{w_n\right\}_{n=1}^{+\infty}$ and a $y\in\mathbb{C}^3$, $\left\|y\right\|=1$ such that $y^{w_n}\rightarrow y$ as $n\rightarrow +\infty$. Just as we did previously, with no loss of generality we may assume $y^w\rightarrow y$ for $w\rightarrow +\infty$. We thus have
\begin{eqnarray}
\lim\limits_{w\to +\infty}\frac{C_{\mathsf{Ad}_{X^w}\circ\Xi_i\circ\mathsf{Ad}_{Y^w}}}{\mathrm{Tr}\left(C_{\mathsf{Ad}_{X^w}\circ\Xi_i\circ\mathsf{Ad}_{Y^w}}\right)}&=&C_{\Sigma}\label{eqlimCPhi}\\
\lim\limits_{w\to +\infty}y^w&=&y\label{eqlimyw}
\end{eqnarray} 
Inequality \eqref{eqRkyCPhi} now trivially follows from \eqref{eqRkyw}, \eqref{eqlimCPhi}, \eqref{eqlimyw} and the continuity of submatrix determinants.  
\end{proof} 
\end{lemma}
As a consequence of the above lemma, the only case where we have not yet proved our assertion is that of $\left<v_2\otimes q_1\right|C_{\Xi_i}\left|v_2\otimes q_1\right>>0$. Because of the arbitrariness of the choice of $i$ and the tensor product side in the above lemma, with no loss of generality we may assume that $\sigma^{w}_2$ tends to zero in a \emph{slowest} way among $\sigma^w_2$, $\sigma^w_3$, $\delta^w_2$, $\delta^w_3$ (otherwise, simply begin the argument with a different singular value). By this, we mean the existence of a subsequence $\left\{w_n\right\}_{n=1}^{+\infty}$ and numbers $\alpha,\beta,\gamma>0$ such that
\begin{eqnarray}
\sigma^{w_n}_3&<&\alpha\sigma^{w_n}_2\label{eqAlpha}\\
\delta^{w_n}_2&<&\beta\sigma^{w_n}_2\label{eqBeta}\\
\delta^{w_n}_3&<&\gamma\sigma^{w_n}_2\label{eqGamma}
\end{eqnarray}
Just like we did earlier a few times, we may neglect terms not in $w_n$ and assume that inequalities \eqref{eqAlpha}-\eqref{eqGamma} hold with $w$ substituted for $w_n$.

Because the denominator of \eqref{eqLimFrac} contains the term $\left(\sigma^w_2\right)^2\left<v_2\otimes q_1\right|C_{\Xi_i}\left|v_2\otimes q_1\right>$, all the terms in the numerator that vanish faster than $\left(\sigma^w_2\right)^2$ result in zero terms in $C_{\Sigma}$ and therefore also in $C_{\Phi}$. This applies to all terms in \eqref{eqLimFrac} with $s=u=2$ and $r\neq 3$ or $t\neq 3$, which clearly implies 
\begin{equation}
\mathrm{rk}\left(\left<u_2\otimes .\right|C_{\Phi}\left|u_2\otimes .\right>\right)=\mathrm{rk}\left(\left<u_2\otimes .\right|C_{\Sigma}\left|u_2\otimes .\right>\right)<2,
\end{equation} 
Hence, the main assertion of this appendix also holds in the case of positive $\left<v_2\otimes q_1\right|C_{\Xi_i}\left|v_2\otimes q_1\right>>0$. 

Note that despite the fact that our proof was carried out for $\Phi$ in $\mathcal{P}_{3,3}$, it is easily generalized to positive maps in $\mathcal{P}_{m,n}$ for $m$ and $n$ arbitrary, however the inequalities \eqref{eqRk1} and \eqref{eqRk2} need to be replaced by the condition that the rank of $\left<x\otimes.\right|C_{\Phi}\left|x\otimes .\right>$ or $\left<.\otimes y\right|C_{\Phi}\left|.\otimes y\right>$ must be smaller than the maximum possible value minus one. 

\section{A class of moduli preserving linear transformations}\label{appModuluses}
In this appendix, we prove that  $n\times n$ matrices $X$ such that 
\begin{equation}
\left|\sum_{l=1}^nX_{kl}e^{i\phi_l}\right|=\left|\sum_{t=1}^nX_{mt}e^{i\phi_t}\right|\forall_{\left\{\phi_l\right\}_{l=1,2,\ldots,n}}\forall_{k,m\in\left\{1,2,\ldots,n\right\}}
\end{equation}
are necessarily singular or they only have a single nonzero entry in each row. To show this, it is enough to prove the following lemma.
\begin{lemma}
Let $y$ and $z$ be vectors in $\mathbb{C}^n$ such that
\begin{equation}\label{eqVecMod}
\left|\sum_{l=1}^ny^le^{i\phi_l}\right|=\left|\sum_{t=1}^nz^te^{i\phi_t}\right|\forall_{\left\{\phi_l\right\}_{l=1,2,\ldots,n}}
\end{equation}
where $y^l$ and $z^t$ are the coefficients of the vectors $y$ and $z$, respectively. Then $y$ and $z$ must either be proportional to each other or each of them must have exactly one nonzero coefficient.
\begin{proof}
We prove the lemma by induction on $n$. For $n=1$, its assertion is obviously true. For $n>1$, let us write the moduli in \eqref{eqVecMod} as
\begin{equation}\label{eqVecInd}
\left|y^1e^{i\phi_1}+\sum_{l=2}^ny^le^{i\phi_l}\right|=\left|z^1e^{i\phi_1}+\sum_{t=2}^nz^te^{i\phi_l}\right|
\end{equation}
A quick thought reveals that there are only four cases where \eqref{eqVecInd} can hold for all possible choices of the $\phi_l$'s. These are:
\begin{enumerate}
\item 
\begin{eqnarray}
y^1&=&0\\
z^1&=&0\\
\forall_{\left\{\phi_l\right\}_{l=2}^n} \left|\sum_{l=2}^ny^le^{i\phi_l}\right|&=&\left|\sum_{l=2}^nz^te^{i\phi_t}\right|
\end{eqnarray}
\item
\begin{eqnarray}
y^1&=&0\\
z^1&\neq&0\\
\exists_{\alpha>0}\forall_{\left\{\phi_l\right\}_{l=2}^n}\left|\sum_{l=2}^ny^le^{i\phi_l}\right|&=&\alpha\label{eqConstAlpha}\\
\forall_{\left\{\phi_t\right\}_{t=2}^n}\left|\sum_{t=2}^nz^te^{i\phi_t}\right|&=&0\label{eqConstZero} 
\end{eqnarray}
\item
\begin{eqnarray}
y^1&\neq&0\\
z^1&=&0\\
\exists_{\beta>0}\forall_{\left\{\phi_t\right\}_{t=2}^n}\left|\sum_{t=2}^nz^te^{i\phi_t}\right|&=&\beta\\
\forall_{\left\{\phi_l\right\}_{l=2}^n}\left|\sum_{l=2}^ny^le^{i\phi_l}\right|&=&0 
\end{eqnarray}
\item
\begin{eqnarray}
y^1&\neq&0\\
z^1&\neq&0\\
\forall_{\left\{\phi_l\right\}_{l=2}^n}\frac{\sum_{l=2}^ny^le^{i\phi_l}}{y^1}&=&\frac{\sum_{t=2}^nz^te^{i\phi_t}}{z^1}
\end{eqnarray}
\end{enumerate}
In case 1), the assertion of the lemma automatically holds by induction. In case 2), equation \eqref{eqConstAlpha} can only hold if exactly one of the numbers $y^2,y^3,\ldots,y^n$ is nonzero. Moreover, equation \eqref{eqConstZero} can only hold if all the numbers $z^2,z^3,\ldots,z^n$ are equal to zero. Altogether, $y$ and $z$ have exactly one nonzero coordinate, which fulfills the assertion of the lemma. Case 3) is symmetric to case 2). In case 4), we must have
\begin{equation}\label{eqProportional}
y^l=\frac{y^1}{z^1}z^l
\end{equation} 
for $l=2,3,\ldots,n$. However, equation \eqref{eqProportional} obviously also holds when $l=1$. In other words, the vectors $y$ and $z$ turn out to be proportional to each other, just as in the assertion of the lemma. 
\end{proof}
\end{lemma}

From the above lemma, the main conclusion of the appendix easily follows if we take as $y$ and $z$ arbitrary two rows of the matrix $X$. In case they turn out to be proportional to each other, the matrix turns out to be singular. Otherwise, there is exactly one nonzero entry in each of these rows. This argument can be repeated for all row pairs in $X$ to yield the main result of this appendix. 

\bibliographystyle{unsrt}
\bibliography{2vs3BoundEntanglement}{}

\end{document}